\documentclass{easychair}

\usepackage[utf8]{inputenc}
\usepackage{lstcoq}

\title{Generating induction principles and subterm relations \\ for inductive types using MetaCoq}
\author{Bohdan Liesnikov\inst{1}\inst{2} \and Marcel Ullrich\inst{1} \and Yannick Forster\inst{1}}
\institute{Saarland University, Saarland Informatics Campus, Saarbrücken, Germany \and IMPRS-CS, Saarland Informatics Campus, Saarbrücken, Germany}

\titlerunning{}
\authorrunning{}

\begin{document}
\maketitle
\pagestyle{empty}

When one defines an inductive type in Coq there are various derivable definitions one frequently uses, like induction principles, subterm relations, equality deciders, countability proofs, alternative representations of constructors, etc.
Some of these can be derived automatically by Coq or Coq plugins.
For instance, Coq derives induction principles automatically:
When one defines a type of balanced trees with leaves labelled by elements of \lstinline{A} as follows
\begin{coq}
  Inductive brtree A : nat -> Type :=
  | Leaf (a : A) : brtree A 0   | Node (n : nat) (l : list (brtree A n)) : brtree A (S n).
\end{coq}
Coq automatically derives the principle
\begin{coq}
  brtree_rect : forall (A : Type) (P : forall n, brtree A n -> Type),
  (forall a, P 0 (Leaf A a)) -> (forall n l, P (S n) (Node A n l)) -> forall n (t : brtree A n), P n t.
\end{coq}
As is well-known, the automatic derivation ignores the nested recursive occurrence of \lstinline{brtree} in the argument \lstinline{l} of the constructor \lstinline{Node}.

Furthermore, the \textsc{Equations}~\cite{sozeau2019equations} plugin for Coq for instance comes with a command
\begin{coq}
  Derive Subterm for brtree.
\end{coq}
which derives the direct subterm relation for \lstinline{brtree} and tries to prove it well-founded -- but again this command ignores the recursive occurrence in \lstinline{Node}.

Tassi's Elpi plugin~\cite{tassi} already covers induction principles for nested inductive types like \lstinline{brtree}.
It would be interesting to directly extend Coq's induction principle mechanism to cover nested inductives as well.
However, writing Coq plugins and extending the Coq source code is at least highly non-trivial.
For non-experts, the OCaml code of both Coq and Coq plugins is hard to access and almost impossible to adapt without the help of experts.

The MetaCoq project~\cite{sozeau:hal-02167423} aims at making the implementation of Coq plugins easier.
Instead of writing OCaml code, one implements a syntax transformation over an inductive type of terms as pure Coq function.
We present three plugins written in this style and hope that the proposed presentation can make writing MetaCoq plugins more accessible.

Our plugins are available at \url{https://github.com/uds-psl/metacoq-examples-coqws}.

\paragraph*{Alternative constructors for types}

When one works with dependent inductive types like \lstinline{brtree} it might be helpful to define
\begin{coq}
  Node_eqs : forall n (l : list (brtree A b)) m, m = S n -> brtee A m
\end{coq}

Writing a syntax transformation on MetaCoq's inductive type of Coq terms transforming the type of \lstinline{Node} to the type of \lstinline{Node_eqs} is straightforward.
In MetaCoq, the type \lstinline{forall x : A. B} for instance is represented by the element \lstinline{tProd (nNamed "x") A B} of the inductive type \lstinline{term}, and bindings are implemented by \lstinline{tRel n} using de Bruijn indices.
We provide a command
\begin{coq}
  MetaCoq Run Derive Generalized Constructor for Node as Node_eqs.
\end{coq}%
applicable to non-mutual inductive types, which derives \lstinline{Node_eqs} via the following function:
\begin{coq}
Fixpoint abstract_eqns (Σ : global_env_ext) (Γ : context) (ty : term) (n : nat) : term :=
  match ty with
  | tProd na A B =>
    tProd na A (abstract_eqns Σ (Γ,,na ↦ A) B 0)
  | tApp L A =>
    let type_of_x := try_infer Σ Γ (lift (2 * n) 0 A) in (* returns type or tRel 0 *)
    let eqn := mkApps tEq [type_of_x; tRel 0; lift (1 + 2 * n) 0 A] in 
    tProd (nNamed "x") type_of_x
    (tImpl eqn (abstract_eqns Σ (Γ,"x" ↦ type_of_x,eqn) L (S n)))
  | B => mkApps B (map (fun m => tRel (1 + 2 * m)) (seq 0 n))
  end.
\end{coq}
\vspace{-5mm}
\paragraph*{Induction principles for nested inductive types}

Similarly, we implement a syntax transformation generating a proof term of an induction principle given a representation of an inductive type.
Using ideas from Tassi's Elpi plugin~\cite{tassi} we support non-mutual nested inductive types by re-using the unary parametricity translation implemented in~\cite{anand2018towards} and provide a command
\begin{coq}
  MetaCoq Run Scheme Induction for brtree.
\end{coq}
which derives the strongest possible induction principle:
\begin{coq}
  brtree_ind_MC : forall p : forall (A : Type) (n : nat), brtree A n -> Type,
  (forall A a, p A 0 (Leaf A a)) ->
  (forall A n l, is_list (brtree A n) (p A n) l -> p A (S n) (Node A n l)) ->
  forall A n t, p A n t
\end{coq}
Note that \lstinline{is_list} is essentially the \lstinline{List.Forall} relation lifted to \lstinline{Type}.

\vspace{-4mm}
\paragraph*{Subterm relation}

Lastly, we implement a syntax transformation which generates the subterm relation for non-mutual inductive types (not covering nested inductives).
For instance,
\begin{coq}
  MetaCoq Run Derive subterm for list.
\end{coq}
derives
\begin{coq}
  Inductive list_direct_subterm : forall A : Type, list A -> list A -> Prop :=
  | cons_subterm0 : forall (A : Type) (a : A) (l : list A), list_direct_subterm A l (a :: l).
\end{coq}

\vspace{-5mm}
\enlargethispage{1.5cm}
\paragraph*{Future work}
It should not be hard to extend the derivation of subterm relations to also cover nested inductive types.
In principle MetaCoq also allows the verification of plugins, for instance by relying on the verified type inference from~\cite{sozeau2019coq} one could prove that the term generated in the first plugin is well-typed, that the type of the induction principle is well-formed, or that the reflexive transitive closure of the subterm relation is well-founded.
We are trying to find the right abstractions to make such proofs feasible.
It would also be interesting to implement automatic countability and finiteness proofs in MetaCoq, like~\cite{amorim} does based on type classes.

\vspace{-5mm}
\bibliography{bib.bib}
\bibliographystyle{plain}

\end{document}